\documentstyle[12pt]{article}
\begin{document}

\title{Chapman's model for ozone concentration: earth's slowing rotation effect in
the atmospheric past}
\author{J. C. Flores$^{\dagger }$ and S. Montecinos$^{\ddagger }$}
\date{$^{\dagger }$Universidad de Tarapac\'a, Departamento de F\'\i sica, Casilla
7-D, Arica, Chile.\\
$^{\ddagger }$Universidad de la Frontera, Departamento de F\'\i sica,
Casilla 54-D, Temuco, Chile.}
\maketitle

Chapman's model for ozone concentration is studied. In this nonlinear model,
the photodissociation coefficients for $O_2$ and $O_3$ are time-depending
due to earth-rotation. From the Kapitsa's method, valid in the high
frequency limit, we find the criterion for the existence of equilibrium
solutions. These solutions are depending on the frequency, and require a
rotation period $T$ which satisfies $T<T_1$ or $T>T_2$. Where the critical
periods $T_1$ and $T_2$, with $T_2>T_1$, are a function of the parameters of
the system (reaction rates and photodissociation coefficients). Conjectures
respect to the retardation of the earth's rotation, due to friction, suggest
that the criterion was not even verified in the atmospheric past.

$$
{} 
$$

Oscillations.

\newpage\ 

\begin{section}*{1.- Introduction}
\end{section}

The dynamics of the ozone layer in the atmosphere has different basic
process like: chemical reactions, photochemical reactions and transport
(diffusion, convection, etc.). In a general point of view, this dynamics is
complex and requires some approximations to be studied. In this sense, we
consider a photochemical model proposed by S. Chapman (Brasseur, 1986;
Chapman, 1930; Wayne, 1991). This model considers a set of reactions between
the oxygen components. Explicitly,

$$
{} 
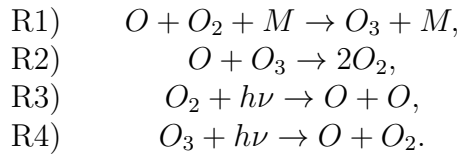
$$

\begin{tabular}{ccccc}
&  &  &  &  \\ 
R1) &  & $O+O_2+M\to O_3+M,$ &  &  \\ 
R2) &  & $O+O_3\to 2O_2,$ &  & \\
R3) &  & $O_2+h\nu \to O+O,$ &  &  \\ 
R4) &  & $O_3+h\nu \to O+O_2.$ &  &  \\ 
\end{tabular}

$$
{} 
$$

In the reaction of the ozone production R1, $M$ denotes any atmospheric
element acting as catalyzer. The reaction R2 denotes the loss of oxygen $O$
and ozone $O_3$ producing molecular oxygen $O_2$. R3 and R4 correspond to
photochemical destruction process related to the solar radiation (symbolized
by $h\nu $).

$$
{} 
$$

The time evolution equations for the constituents consider the above
reactions for variable concentration assuming the concentration of $O_2$
being stationary. Let $X$ and $Y$ be the concentration of $O$ and $O_3$
respectively. Then, the Chapman's model (see for instance Brasseur, 1986;
Montecinos, 1998; 1999) considers the time evolution equations for the
concentrations given by

\begin{equation}
\frac{dX}{dt}=J_1+J_2Y-\left( k_1+k_2Y\right) X, 
\end{equation}

\begin{equation}
\frac{dY}{dt}=k_1X-\left( J_2+k_2X\right) Y,
\end{equation}
where, on the right hand, the positive terms are production rates, and the
negative ones are loss rates. In the nonlinear system (1,2), the quantities $%
J_1$ and $J_2$ are related to the reactions R3 and R4 and correspond to the
photodissociation of $O_2$ and $O_3$ respectively. The important fact is
that they dependent on the sun's radiation, and then, they are periodic in
time with a period $T=24$ hours. In this paper, and by simplicity, we assume 
\begin{equation}
J_i(t)=J_i^o(1-\cos \omega t),~~~i=1,2
\end{equation}
where $\omega =2\pi /T$ , and $J_i^o$ are positive constants. On the other
hand, the positive constants $k_1$ and $k_2$ in (1,2), are temperature
dependent. Also they are dependent on the $O_2$ concentration, and related
to the reaction velocity in R1 and R2 respectively (DeMore, 1994).

$$
{} 
$$

In this paper we propose an analytical study of the nonlinear model (1-3).
In a general point of view, the study of this system is a difficult task,
nevertheless, some interesting results can be find in the high frequency ($%
\omega $) limit. In fact, we use a method proposed originally by Kapitsa
(Landau, 1982) for a mechanical particle in a field with rapid temporal
oscillations in the parameters and nonlinear terms. We find explicitly the
solutions of the systems (1-3) in the high frequency regime (19,20). They
are the equilibrium solution for the time-averaged concentration.
Calculations tell us that no solution exist in some frequency range. The
study of the behavior of this model, for different frequencies, has a
physical interest because earth period of rotation changes with the
geological age. It is a known fact that the rotation of earth is gradually
slowing down by friction. In fact, the estimated rotation velocity
diminishes by 4.4 hours every billon of years (Shu, 1982). 
$$
{} 
$$

For explicit calculations, we assume the following values for the
parameters: 
\begin{equation}
J_1^o\sim 10^7[1/s];~~~~J_2^o\sim 10^{-3}[1/s];~~~~k_1\sim
10[1/s];~~~~k_2\sim 2.5\times 10^{-15}[1/s], 
\end{equation}
corresponding to the values for the ozone layer altitude at, more or less,
35 km. 
$$
{} 
$$

Finally we note that the case of small frequency ($\omega \rightarrow 0$)
can be solved. In fact, here the parameters $J_i(t)$ evolve slowly with time
and then, they can be assumed as constant in the integration process of
(1,2). So the solutions corresponding to the `fixed point' ($\frac{dx}{dt}%
\sim \frac{dy}{dt}\sim 0$, for definitions see (Seydel, 1988; Wio, 1997)) are

\begin{equation}
Y=-\frac{J_1^o}{4J_2^o}+\sqrt{\left( \frac{J_1^o}{4J_2^o}\right) ^2+ \frac{%
k_1J_1^o}{2k_2J_2^o}}, 
\end{equation}

\begin{equation}
X=\frac Y{k_1-k_2Y}J_2\left( t\right) . 
\end{equation}

Namely, in this approximation, the variable $Y$ is constant and $X$ varies
linearly with the dissociation coefficient. Moreover, the solutions (5,6)
are consistent with the numerical solution in the stratosphere (Fabian,
1982; Montecinos, 1996). We note that, from (5), it is easy to show that the
variable $Y$ satisfies $Y\leq \frac{k_1}{k_2}.$ In fact, it is a general
bound for the solution of the systems (1,2) (Montecinos, 2000).

$$
{} 
$$

\begin{section}*{2.- The method of Kapitsa }
\end{section}

As said before, we shall study the system (1-3) in the high frequency
regime. High frequency means here small period of oscillations $T$ respect
to the relaxation-time $T_R$ for the slow variables.

$$
{} 
$$

Assume the separation of the concentrations $X$ and $Y$ in a slow temporal
variation ($x$ and $y$) and other fast ($\varepsilon $ and $\eta ,$
respectively). Namely,

\begin{equation}
X=x+\varepsilon ;\quad Y=y+\eta , 
\end{equation}
where the fast variables are periodic with temporal average zero, namely,

\begin{equation}
\left\langle \varepsilon \right\rangle _T=\left\langle \eta \right\rangle
_T=0. 
\end{equation}
The equation (1) can be re-written like

$$
\frac{dx}{dt}+\frac{d\varepsilon }{dt}=J_1^o+J_2^oy-\left( k_1+k_2y\right)
x-\left( J_1^o+J_2^oy\right) \cos \omega t+ 
$$
\begin{equation}
\left( J_1^o-J_2^o\cos \omega t-k_2x\right) \eta -\left( k_1+k_2y\right)
\varepsilon -k_2\varepsilon \eta , 
\end{equation}
and equation (2) becomes

$$
\frac{dy}{dt}+\frac{d\eta }{dt}=k_1x-\left( J_2^o+k_2x\right) y+J_2^oy\cos
\omega t- 
$$
\begin{equation}
\left( J_2^o+k_2x-J_2^o\cos \omega t\right) \eta +\left( k_1-k_2y\right)
\varepsilon -k_2\varepsilon \eta . 
\end{equation}

On the other hand, the fast variables are only related to rapid oscillation
(Landau, 1982). In this way, from the above expression (9,10), they are
assumed to be a solution to the differential equations:

\begin{equation}
\frac{d\varepsilon }{dt}=-\left( J_1^o+J_2^oy\right) \cos \omega t;\quad 
\frac{d\eta }{dt}=J_2^oy\cos \omega t. 
\end{equation}

At this point a remark becomes necessary. The expression (7), complemented
with the above equations (11), defines a change of variables without
approximations. Nevertheless, the differential equations (11) are suggested
by the direct oscillatory term in (9) and (10). Kapitsa's method consider
the equations for $\frac{d\varepsilon }{dt}$ and $\frac{d\eta }{dt}$ as
approximated.

$$
{} 
$$

In one period, the slow variables are essentially constants and the fast
have zero average (8), then, the time average of the equation (9) becomes.

\begin{equation}
\frac{dx}{dt}=J_1^o+J_2^oy-\left( k_1+k_2y\right) x-J_2^o\left\langle \eta
\cos \omega t\right\rangle _T-k_2\left\langle \varepsilon \eta \right\rangle
_T, 
\end{equation}
and, for equation (10), we obtain 
\begin{equation}
\frac{dy}{dt}=k_1x-\left( J_2^o+k_2x\right) y+J_2^o\left\langle \eta \cos
\omega t\right\rangle _T-k_2\left\langle \varepsilon \eta \right\rangle _T. 
\end{equation}
Nevertheless, since equations (11) can be solved exactly, 
\begin{equation}
\varepsilon (t)=-\frac 1\omega \left( J_1^o+J_2^oy\right) \sin \omega
t;\quad \eta (t)=\frac{J_2^o}\omega y\sin \omega t, 
\end{equation}
the evolution equations for the slow variables become

\begin{equation}
\frac{dx}{dt}=J_1^o+J_2^oy-\left( k_1+k_2y\right) x+\frac{J_2^ok_2}{2\omega
^2}y\left( J_1^o+J_2^oy\right) , 
\end{equation}
and 
\begin{equation}
\frac{dy}{dt}=k_1x-\left( J_2^o+k_2x\right) y+\frac{J_2^ok_2}{2\omega ^2}%
y\left( J_1^o+J_2^oy\right) . 
\end{equation}

This set of equations are the basis for our analytical results. They are
restricted to the high frequency approximation. This approximation becomes
given by the `expansion' in $1/\omega ^2$ related to the last term in
(15,16). Remark that it is an autonomous nonlinear systems and then, without
the explicit temporal dependence. This transformation, from a set of
equations with time-periodic parameters, to other autonomous, is related to
the Kapitsa original ideas (Landau, 1982).

$$
{} 
$$

In a general frame of work, the system (15,16) is complex. Moreover, the
approximation of high frequency is valid when the relaxation time $T_R,$ of
the equations (15,16), is bigger than $2\pi /\omega $. This comparison is a
difficult task, nevertheless, the case {\bf \ $J_2^o=0$ }can be solved
exactly to estimate the validity of the approximation.{\bf \ }It corresponds
formally to eliminate the dissociation of $O_3$. The asymptotic solution of
the non-autonomous system (1-3) is (Montecinos, 2000).

\begin{equation}
X=\frac{J_1^o}{2k_1}-\frac{J_1^o}{\sqrt{4k_1^2+\omega ^2}}\cos (\omega
t-\phi ),\qquad Y=\frac{k_1}{k_2}, 
\end{equation}
where the phase $\phi $ is given by the relation: $\tan \phi =\omega /2k_1$.
On the other hand, combining the equations (15,16) with (7), in the high
frequency approximation we found that the equilibrium solution given by the
Kapitsa's method is: 
\begin{equation}
X=\frac{J_1^o}{2k_1}-\frac{J_1^o}\omega \sin \omega t;\qquad Y=\frac{k_1}{k_2%
}. 
\end{equation}
It is direct to show that the exact solution (17) reduces to (18) in the
high frequency limit. Moreover, the relaxation time $T_R$ can be calculated
here analytically. It is given by $T_R=1/2k_1$. So, we expect that the high
frequency approximation (15,16) is valid when $4\pi k_1$ $\ll \omega $.

$$
{} 
$$

\begin{section}*{3.- Existence of  equilibrium solutions }
\end{section}

In this section we are concerned with the fixed point solution (Seydel,
1988; Wio, 1997) of the autonomous set (15,16). This system have an
equilibrium point $(x_o,y_o)$, defined by $\frac{dx}{dt}=\frac{dy}{dt}=0,$
and given by the solution of the equations:

\begin{equation}
k_2J_2^o\left( 2-\frac{k_1J_2^o}{\omega ^2}\right) y_o^2+k_2J_1^o\left(1-%
\frac{k_1J_2^o}{\omega ^2}\right) y_o-k_1J_1^o=0, 
\end{equation}
and

\begin{equation}
x_o=\frac{\left( J_1^o+J_2^oy_o\right) \left( 1+\frac{k_2J_2^o}{2\omega ^2}%
y_o\right) }{\left( k_1+k_2y_o\right) }. 
\end{equation}

Equations (19) and (20) define the homogeneous equilibrium solution of the
autonomous system (15) and (16) and then, with (7) and (14), we have the
solution of the systems (1,2) in the high frequency regime.

$$
{} 
$$

The existence of real solutions, for the second degree equation (19),
requires the inequality

\begin{equation}
\frac 1{\omega ^4}-\frac 4{k_2J_1^o}\left( 1+\frac{k_2J_1^o}{2k_1J_2^o}%
\right) \frac 1{\omega ^2}+\frac 8{k_1k_2J_1^oJ_2^o}\left( 1+ \frac{k_2J_1^o%
}{8k_1J_2^o}\right) \geq 0, 
\end{equation}
which corresponds to an inequality of second degree for $1/\omega ^2$.
Namely, there is no equilibrium solution of (15,16) if and only if,

\begin{equation}
\frac 2{k_2J_1^o}+\frac 1{k_1J_2^o}-\frac 2{k_2J_1^o}\sqrt{1-\frac{k_2J_1^o}{%
k_1J_2^o}}\leq \frac 1{\omega ^2}\leq \frac 2{k_2J_1^o}+\frac
1{k_1J_2^o}+\frac 2{k_2J_1^o}\sqrt{1-\frac{k_2J_1^o}{k_1J_2^o}}. 
\end{equation}

$$
{} 
$$

If we assume the condition

\begin{equation}
\frac{k_2J_1^o}{k_1J_2^o}\ll 1, 
\end{equation}
valid for the parameters (4) of section 1, the inequality (22) can be
re-written for the period $T$. In fact, there is no equilibrium solution of
the system (15,16) when

\begin{equation}
T_1\leq T\leq T_2\quad (no-solution), 
\end{equation}
where

\begin{equation}
T_1=2\pi \sqrt{\frac 2{k_1J_2^o}}; \quad T_2=4\pi \sqrt{\frac 1{k_2J_1^o}}. 
\end{equation}

$$
{} 
$$

\begin{section}*{4.- Earth's slowing rotation and the existence of solution}
\end{section}

It is interesting that the inequality (24) gives a region were no solution
exist. Here we must take care because no oscillating solution like (7)
exist. In fact, (24) splits the $\omega -$space parameter in three regions:
(i) The region defined by $T<T_1$ where a real positive solution ($y_o>0$)
of equation (15) exist, with a negative one ($y_o<0$) . (ii) The region
defined by (24), where no solution of (15) exist. (iii) The region defined
by $T>T_2$ where solutions are negative ($y_o<0$). From (5,6), we known that
in the slow frequency limit (region (iii)) a real positive solution exist.
Then, Kapitsa's method does not work well in this region, nevertheless, at
least it says that an oscillating solution exist.

$$
{} 
$$

At this point we can formulate the following question: since the
earth-rotation has diminished by friction, how has the change in rotation
affected the existence of the ozone layer ?. This question seems appropriate
because the Kapitsa's method tells us that the frequency of rotation and the
ozone concentration are related. Using the parameter values (4), of section
1, we can estimate the critical period (24) : $T_1\sim 0.02$ hours and $%
T_2\sim 22$ hours. Is interesting that the actual period $T=24$ hours, is in
the region of permitted solution ($T>T_2$). Moreover this is suggestive:
from the retardation of earth rotation velocity data (4.4 hours/billon of
years, (Shu, 1982)), a simple calculation tells us that before $\frac{24-22}{%
4.4}\sim 0.46$ billons years no solution existed because we were in the
region (24). This is a surprising estimation if we consider that actually
the ozone layer is believed to have been in existence $0.7$ billon years
(Graedel, 1993).

$$
{} 
$$

\begin{section}*{5.- Conclusions }
\end{section}

We have considered the Chapman's model for ozone production (1,2). In this
nonlinear model, the parameters related to photodissociation are periodic in
time (3). We were interested at the analytical study of this model by using
the high frequency approximation, due to Kapitsa. Namely, we have considered
the autonomous system (15,16), depending on the frequency, for the averaged
variable concentrations. The existence of equilibrium solutions (fixed
points (19,20)) is depending on the frequency. In fact, there are two
critical period $T_1$ and $T_2$ so that for $T_1<T<T_2$ there is no
equilibrium solution (24).

$$
{} 
$$

The values for the parameters (4), in section 1, give the condition of
no-existence (24): $0.02$ hours $\leq T\leq 22$ hours, and then compatible
with the actual earth's period of rotation, and existence of the ozone
layer. Moreover, considering the earth's slowing rotation motion due to
friction (4.4 hours every billon of years, (Shu, 1982)), we estimate that
the ozone existence condition is verified after $\frac{24-22}{4.4}\sim 0.46$
billon years (section 4). This is a good estimation if we consider the
simplicity of the autonomous model given by equations (15,16). The age of
the ozone layer is 0.7 billons of years (Graedel, 1993)).

$$
{} 
$$

Before to ending a remark, equations (15) and (16) are very adequate to the
study of diffusion process, which was neglected in the original equations
(1) and (2). In fact, because they are not time depending, when we add
spatial diffusion terms $D\frac{d^2}{dx^2}X$ and $D\frac{d^2}{dx^2}Y$, they
become similar to reaction-diffusion-equations.

$$
{} 
$$

{\bf Acknowledgments:} This work was possible thanks to Project UTA-Mayor
4725 (Universidad de Tarapac\'a). Useful discussion with professor H. Wio,
D. Walgraef (visits supported by the FDI-UTA and PELICAN Projects) and M.
Pedreros, are acknowledged.

$$
{} 
$$


\begin{thebibliography}{99}
\bibitem{Brass}  Brasseur G. \& Solomon S., 1986: {\it Aeronomy of the
Middle Atmosphere}, D. Reidel pub. comp., Holland.

\bibitem{chapman}  Chapman S., 1930: {\it A theory of upper--atmosphere ozone%
}, Mem. Roy. Meteorol. Soc. {\bf 3}, 103.

\bibitem{jpl94}  DeMore , W. B., {\it et al}, 1994: {\it Chemical Kinetics
and Photochemical Data for Use in Stratospheric Modeling}, Eval. {\bf 11},
Natl. Aeronaut. and Space Admin., Jet Propul.Lab., Pasadena.

\bibitem{fabian}  Fabian P., Pyle J. A. and Wells J. R., 1982: {\it Diurnal
Variations of Minor Constituents in the Stratosphere Modeled as a Function
of Latitude and Season}, J. Geophys. Res. {\bf 87}, 4981.

\bibitem{}  Graedel T. E. and Crutzen P. J., 1993: {\it Atmospheric Changes:
An Earth Systems Perpective}, W. H. Freeman Company, N. Y.

\bibitem{landau}  Landau L. and Lifchitz E., 1982: {\it Physique Theorique,
Vol. 1, Mecanique}, Edition Mir.

\bibitem{tesis}  Montecinos S., 1996: {\it Reaktionskinetische
Photochemische Modellierung de Ozonkonzentration der Mesosph\"are},
Dissertation, Papier Flieger Verlag, Clausthal--Zellerfeld (unpublished).

\bibitem{sondoeb}  Montecinos S., Doebner H. D., 1998: {\it Dynamical
systems based on a mesospheric photochemical model}, Phys. Lett. {\bf 241A},
269.

\bibitem{syc}  Montecinos S. , Flores J. C. , 2000: {\it Soluciones anal\'\i
ticas del modelo atmosf\'erico de Chapman}, submitted.

\bibitem{}  Montecinos S., Felmer P., 1999: {\it Multiplicity and Stability
of Solutions for Chapman and Mesospheric Photochemical Model}, J. Geophys.
Res. {\bf 104}, 11799.

\bibitem{}  Seydel R., 1988: {\it From Equilibrium to Chaos}, Elseiver, N. Y.

\bibitem{}  Shu F. H., 1982: {\it The Physical Universe}, University Science
Books, Mill Valley, California.

\bibitem{wayne}  Wayne R. P., 1991: {\it Chemistry of the Atmosphere},
Oxford University Press Inc., N. Y.

\bibitem{}  Wio H., 1997: {\it Computational Physics} (P. L. Garrido and J.
Marro, Eds.), Springer Verlag.
\end{thebibliography}
\end{document}